# On generalized scaling laws with continuously varying exponents

**Lionel Sittler and Haye Hinrichsen**

Theoretische Physik, Fachbereich 8, Universität Wuppertal, 42097 Wuppertal, Germany

**Abstract**
Many physical systems share the property of scale invariance. Most of them show ordinary power-law scaling, where quantities can be expressed as a leading power law times a scaling function which depends on scaling-invariant ratios of the parameters. However, some systems do not obey power-law scaling, instead there is numerical evidence for a logarithmic scaling form, in which the scaling function depends on ratios of the *logarithms* of the parameters. Based on previous ideas by Tang we propose that this type of logarithmic scaling can be explained by a concept of local scaling invariance with continuously varying exponents. The functional dependence of the exponents is constrained by a homomorphism which can be expressed as a set of partial differential equations. Solving these equations we obtain logarithmic scaling as a special case. The other solutions lead to scaling forms where logarithmic and power-law scaling are mixed.

PACS numbers: 05.70.Jk, 64.60.Fr, 05.50.+q

## 1. Introduction

Various physical systems in equilibrium and non-equilibrium statistical physics exhibit scaling invariance, i.e. their behaviour does not change under rescaling of space and time combined with an appropriate rescaling of the observables and the control parameters [1, 2]. Typically scaling invariance implies that quantities such as, e.g., order parameters are quasi-homogeneous functions $f$ which obey a functional equation of the form [3]

$$f(a^{v_1}x_1, \ldots, a^{v_n}x_n) = af(x_1, \ldots, x_n). \qquad (1)$$

Here $a$ is a dilatation parameter, $x_i$ can be the time, space, system size or any other control parameter of the system and the $v_i$ are certain exponents. These exponents are constant, i.e. they do not depend on the variables $x_i$. Performing a projective map, one can define a new set of $n-1$ homogeneous coordinates

$$u_i = \frac{x_i}{x_n^{v_i/v_n}} \qquad \text{for} \quad i=1,\ldots,n-1. \qquad (2)$$

Using these coordinates and choosing $a = x_n^{-1/v_n}$ in equation (1), the function $f$ can be expressed as

$$f(x_1, \ldots, x_n) = x_n^\lambda g(u_1, \ldots, u_{n-1}) \tag{3}$$

where $g$ is a scaling function depending on $n-1$ arguments and $\lambda = 1/v_n$ is a scaling exponent. In numerical studies this scaling form is used to generate a data collapse by plotting $x_n^{-\lambda} f(x_1, \ldots, x_n)$ versus $u_1, \ldots, u_{n-1}$.

The scaling form (3) is valid for a large variety of critical phenomena and systems with self-similar properties. However, there are exceptional cases where simple power-law scaling does not hold, instead there is numerical evidence that the *logarithms* of the parameters can be used to produce a data collapse. Examples include certain self-organized critical sandpile models [4, 5], DLA-related growth processes [6, 8] and even experiments in turbulence [7]. In all these studies, using the notation introduced above, the numerical data seem to obey a logarithmic scaling law of the form

$$\frac{\ln f(x_1, \ldots, x_n)}{\ln x_n} = G\left(\frac{\ln x_1}{\ln x_n}, \ldots, \frac{\ln x_{n-1}}{\ln x_n}\right). \tag{4}$$

Roughly speaking, this scaling form differs from the previous one in so far as the function $f$ and the variables $x_i$ have been replaced by their logarithms. As in equation (3), this logarithmic scaling law reduces the number of independent variables from $n$ to $n-1$. However, so far it is not yet clear to what extent equation (4) can be substantiated theoretically and how it is related to ordinary power-law scaling, especially if both types of scaling coexist in the same model.

In an unpublished preprint from 1989, Tang presented a beautiful idea how the two scaling concepts in equations (3) and (4) can be linked [9]. The main idea goes back to earlier works by Coniglio and Marinaro [10], who suggested considering continuously varying exponents, which depend on the parameters $x_1, \ldots, x_n$. However, their functional dependence is not arbitrary, instead they are strongly constrained by a group homomorphism, as will be explained below. Physically these constraints imply that such a generalized scaling transformation with continuously varying exponents still 'looks like' an ordinary scaling transformation in any co-moving infinitesimal volume of the parameter space spanned by the $x_i$. In this way we introduce a new concept of *local scaling invariance*.

The purpose of the present work is to investigate the concept of local scaling invariance and its implications in more detail. To this end we first summarize the essential ideas of Tang's unpublished paper in section 2. However, in contrast to Tang's work, which was restricted to the case $n = 2$ and using a notation adapted to finite-size scaling of avalanches in SOC, it was our aim to present the theory as general as possible, formulating it for $n$ variables, symmetrizing the expressions and using a compact vector notation.

In section 3 we turn to the solutions of the partial differential equations for the critical exponents. Introducing projective coordinates, which provide a better insight into the structure of the problem, we classify the possible solutions. In addition to ordinary and logarithmic scaling forms we find further non-trivial solutions, e.g., mixed scaling forms where logarithmic and power-law behaviour coexists. Finally the concluding section discusses the obtained results.

## 2. Concept of local scaling invariance

Starting point of the multiscaling ansatz is to replace the constant exponents $v_i$ by continuously varying exponents which depend on the variables $x_i$. In analogy to equation (1) a generalized scaling transformation $\tau(a)$ is then defined by

$$\tau(a): \quad f(a^{v_1(x_1,\ldots,x_n)} x_1, \ldots, a^{v_n(x_1,\ldots,x_n)} x_n) = af(x_1, \ldots, x_n) \tag{5}$$

where $a$ is again a dilatation parameter. However, the exponents $v_i(x_1, \ldots, x_n)$ cannot be chosen freely, rather their functional dependence is constrained by a *group homomorphism*. This homomorphism links the concepts of generalized and ordinary scaling by requiring that two subsequent dilatations by the factors $a$ and $b$ are equivalent to a single dilatation by the factor $ab$:

$$\tau(ab) = \tau(a)\tau(b). \tag{6}$$

In other words, if we apply equation (5) for $\tau(b)$ and thereafter for $\tau(a)$, i.e.

$$abf(x_1, \ldots, x_n) = bf(a^{v_1(x_1,\ldots,x_n)}x_1, \ldots, a^{v_n(x_1,\ldots,x_n)}x_n)$$
$$= f\left(b^{v_1(a^{v_1}x_1,\ldots,a^{v_n}x_n)}a^{v_1(x_1,\ldots,x_n)}x_1, \ldots, b^{v_n(a^{v_1}x_1,\ldots,a^{v_n}x_n)}a^{v_n(x_1,\ldots,x_n)}x_n\right) \tag{7}$$

the result should be the same as if we had transformed the system in one step by $\tau(ab)$:

$$abf(x, \ldots, x_n) = f((ab)^{v_1(x_1,\ldots,x_n)}x_1, \ldots, (ab)^{v_n(x_1,\ldots,x_n)}x_n). \tag{8}$$

Comparing equations (7) and (8) we get

$$v_i(x_1, \ldots, x_n) = v_i(a^{v_1}x_1, \ldots, a^{v_n}x_n) \qquad \text{for} \quad i = 1, \ldots, n \tag{9}$$

i.e. the exponents themselves have to be quasi-homogeneous functions. Considering an infinitesimal transformation $a = 1 + \epsilon$ and expanding equation (9) to the first order in $\epsilon$, we find that the exponents $v_i$ satisfy the partial differential equations

$$\left(\sum_{j=1}^{n} v_j x_j \frac{\partial}{\partial x_j}\right) v_i = 0 \qquad i = 1, \ldots, n. \tag{10}$$

Introducing logarithmic coordinates $X_i = \ln|x_i|$ and the vector notation $\mathbf{X} = (X_1, \ldots, X_n)$, $\mathbf{v} = (v_1, \ldots, v_n)$ and $\nabla = (\partial_{X_1}, \ldots, \partial_{X_n})$, equation (10) can be written in the compact form

$$(\mathbf{v} \cdot \nabla)\mathbf{v} = \mathbf{0}. \tag{11}$$

We note that the left-hand side of this equation has the same form as the convective term in the fundamental equation of fluid mechanics, see [11].

Let us now turn to the scaling function $f$. As the exponents $v_i$, the function $f$ has to fulfil a partial differential equation. Considering an infinitesimal transformation $a = 1 + \epsilon$ in equation (5), we obtain to the first order in $\epsilon$

$$f = \sum_{j=1}^{n} v_j x_j \frac{\partial f}{\partial x_j}. \tag{12}$$

Using again logarithmic variables $X_i = \ln|x_i|$ and defining

$$F(X_1, \ldots, X_n) = \ln|f(x_1, \ldots, x_n)|. \tag{13}$$

Equation (12) acquires the simple form

$$\mathbf{v} \cdot \nabla F = 1. \tag{14}$$

This completes the framework of multiscaling. The remaining mathematical task is to solve PDEs (11) for $\mathbf{v}$. Once a solution is found, the corresponding scaling form can be obtained by solving equation (14).

## 3. Solution of the differential equations

*3.1. General solution*

The physical meaning of the partial differential equations (11) can be described as follows. The vector field $\mathbf{v}(\mathbf{X})$ generates trajectories $\mathbf{X}(t)$ as solutions of the equation $\frac{d\mathbf{X}(t)}{dt} = \mathbf{v}$ parametrized by $t$, on which the system moves in the parameter space spanned by the $X_i$ when it undergoes a scaling transformation. Plugging the relation

$$\frac{\partial v_j}{\partial X_i} = \frac{\partial \dot{X}_j}{\partial X_i} = \frac{\ddot{X}_j}{\dot{X}_i} \tag{15}$$

into equation (11), we get

$$\dot{\mathbf{v}} = \ddot{\mathbf{X}} = \mathbf{0} \tag{16}$$

which means that the vector field $\mathbf{v}$ is *constant* along its own trajectories. Consequently, the trajectories have to be straight lines in the space spanned by the $X_i$. Therefore, we have to find a suitable set of coordinates where one component of the vector field is constant. To this end we introduce homogeneous coordinates [3]

$$U_i = \frac{X_i}{X_n} \qquad \text{for} \quad i = 1, \ldots, n-1. \tag{17}$$

We note that for $X_n = 0$ the $U_i$ are undefined. Setting formally $U_n = X_n$, the partial derivatives in the new coordinates are given by

$$\frac{\partial}{\partial X_i} = \frac{1}{U_n} \frac{\partial}{\partial U_i} \qquad \frac{\partial}{\partial X_n} = \frac{\partial}{\partial U_n} - \sum_{j=1}^{n-1} \frac{U_j}{U_n} \frac{\partial}{\partial U_j} \qquad (i = 1, \ldots, n-1). \tag{18}$$

Therefore, the convective operator $(\mathbf{v} \cdot \nabla)$ in equation (11), which acts on each component of $\mathbf{v}$, can be expressed as

$$\mathbf{v} \cdot \nabla = \sum_{j=1}^{n} v_j \frac{\partial}{\partial X_j} = v_n \frac{\partial}{\partial U_n} + \frac{1}{U_n} \sum_{j=1}^{n-1} (v_j - U_j v_n) \frac{\partial}{\partial U_j}. \tag{19}$$

The purpose of the projective transformation (17) is to eliminate one degree of freedom, i.e. the vector field $\mathbf{v}$ should not depend on $U_n$, which means that $\frac{\partial}{\partial U_n}$ vanishes identically in equation (19). Hence we arrive at the equation

$$\mathbf{v} \cdot \nabla v_i = \frac{1}{U_n} \sum_{j=1}^{n-1} (v_j - U_j v_n) \frac{\partial v_i}{\partial U_j} = 0 \qquad \text{for} \quad i = 1, \ldots, n. \tag{20}$$

This equation is satisfied if all terms of the sum vanish identically. This requires that for a given index $j$ either $v_j = U_j v_n$ or $\frac{\partial v_i}{\partial U_j} = 0$. Without loss of generality we can rearrange the indices in such a way that for $j = 1, \ldots, m$ the first condition is satisfied, where $m$ is some integer number in the range $0 \leqslant m < n$. This means that all components of $\mathbf{v}$ are functions of $U_1, \ldots, U_m$. The functions $v_{m+1}(U_1, \ldots, U_m), \ldots, v_n(U_1, \ldots, U_m)$ can be chosen freely, while the functions $v_1(U_1, \ldots, U_m), \ldots, v_m(U_1, \ldots, U_m)$ are related to $v_n$ by (using the original coordinates)

$$v_j = \frac{X_j}{X_n} v_n \left( \frac{X_1}{X_n}, \ldots, \frac{X_m}{X_n} \right) \qquad \text{for} \quad j = 1, \ldots, m. \tag{21}$$

This relation leads to the remarkable result

$$\frac{v_i}{v_j} = \frac{X_i}{X_j} \qquad \text{for} \quad i, j = 1, \ldots, m. \tag{22}$$

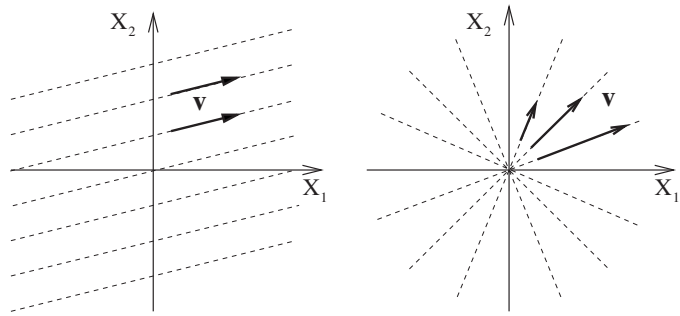

**Figure 1.** Vector field **v** in the space spanned by the logarithmic coordinates $X_1$ and $X_2$. Left: ordinary scaling with constant vector field **v**. Right: generalized scaling according to equation (23). The length of the vectors is determined by the function $h(X_1/X_2)$.

i.e. the ratio of two of the first $m$ components of **v** is equal to the ratio of the corresponding coordinates of **X**.

The choice of a particular projective map in equation (17) is arbitrary. However, the differential equations (11) are invariant under 'translations' $\mathbf{X} \to \mathbf{X} + \Delta\mathbf{X}$ and 'rotations' $\mathbf{X} \to \mathcal{R}\mathbf{X}$. Taking these symmetries into account, equation (21) is the most general solution of equation (11).

Two special cases play an important role. For $m = 0$ all components of **v** are constant so that ordinary power-law scaling is recovered. For $m = n - 1$ only one function can be chosen freely. In this case the solution reads

$$v_i = \frac{X_i}{X_n} h\left(\frac{X_1}{X_n}, \ldots, \frac{X_{n-1}}{X_n}\right) \tag{23}$$

where $h \equiv v_n$ is a scaling function which determines the length of the vectors **v** but not their direction.

### 3.2. Logarithmic scaling in two variables

Let us now consider the case $n = 2, m = 1$, where $f$ depends on two parameters $x_1$ and $x_2$. In this case the exponents are given by

$$\mathbf{v}(\mathbf{X}) = \frac{\mathbf{X}}{X_2} h(X_1/X_2) \tag{24}$$

and describe concentric trajectories originating in the special point $\mathbf{X} = \mathbf{0}$, as shown in the right panel in figure 1. Obviously, the vector field **v** is constant along each of the trajectories while the length of the vector **v** associated with a given trajectory is determined by the scaling function $h$.

Next, we have to solve the equation $\mathbf{v} \cdot \nabla F = 1$, which can be written as

$$h(X_1/X_2)\left(\frac{X_1}{X_2}\frac{\partial}{\partial X_1} + \frac{\partial}{\partial X_2}\right) F(X_1, X_2) = 1. \tag{25}$$

Using the projective map the solution reads

$$F(X_1, X_2) = \frac{X_2}{h(X_1/X_2)} + s(X_1/X_2) \tag{26}$$

where $s(X_1/X_2)$ is an individual offset for each trajectory. Setting $s = 0$ and identifying $h^{-1}$ with $G$ we obtain $F = X_2 G(X_1/X_2)$ or, in the original variables,

$$\frac{\ln f(x_1, x_2)}{\ln x_2} = G\left(\frac{\ln x_1}{\ln x_2}\right) \tag{27}$$

which is the logarithmic scaling form (4) in the case of two variables. This scaling form has been successfully used in various problems [4–8], although in most cases the data collapses were found on a purely empirical basis. The present approach suggests that local scaling invariance may be the physical origin of the observed logarithmic scaling.

### 3.3. Mixed scaling in two variables

The solution for $n = 2$ and $m = 1$ in equation (26) includes not only logarithmic scaling but also other interesting scaling forms where algebraic and logarithmic scaling are mixed. To this end we set

$$h(z) = \frac{1}{\lambda z} \tag{28}$$

where $\lambda$ is a constant. In this case the exponents are given by

$$v_1 = \frac{1}{\lambda} \qquad v_2 = \frac{X_2}{\lambda X_1} \tag{29}$$

and we obtain

$$F(X_1, X_2) = \lambda X_1 + s\left(\frac{X_1}{X_2}\right). \tag{30}$$

In the original coordinates this scaling form reads

$$f = x_1^\lambda s\left(\frac{\ln x_1}{\ln x_2}\right) \tag{31}$$

where the shift $s$ now plays the role of a scaling function. We call it a mixed form of scaling since it combines power-law scaling in the prefactor with logarithmic scaling in the argument.

### 3.4. Restriction on other possible scaling forms

The constraints (12) reduce the possibilities of data collapse. We will show that the homomorphism does not allow all types of data collapses. Suppose that for a given system we find a data collapse based on the scaling form

$$\frac{\ln f}{(\ln x_2)^{\gamma_1}} = k\left(\frac{\ln x_1}{(\ln x_2)^{\gamma_2}}\right) \tag{32}$$

where $\gamma_1$ and $\gamma_2$ are constant exponents. This would correspond to the original scaling form in which all parameters have been replaced by their logarithms. We assume that the function $k$ can be expanded as a power series:

$$\frac{F}{(X_2)^{\gamma_1}} = \sum_n a_n \left(\frac{X_1}{(X_2)^{\gamma_2}}\right)^n. \tag{33}$$

Applying the scaling transformation

$$f \to af \qquad x_1 \to a^{v_1} x_1 \qquad x_2 \to a^{v_2} x_2$$

with non-constant exponents $v_1$ and $v_2$ on both sides of equation (33), we get

$$\frac{F}{X_2} \frac{1 + \frac{\ln a}{F}}{\left(1 + \frac{v_2 \ln a}{X_2}\right)^{\gamma_1}} = \sum_n a_n \left(\frac{X_1}{X_2^{\gamma_2}}\right)^n \left(\frac{1 + \frac{v_1 \ln a}{X_1}}{\left(1 + \frac{v_2 \ln a}{X_2}\right)^{\gamma_2}}\right)^n. \tag{34}$$

Therefore, the exponents **v** satisfy

$$\left(1 + \frac{v_1 \ln a}{X_1}\right)^{1/\gamma_2} = \left(1 + \frac{\ln a}{F}\right)^{1/\gamma_1} = 1 + \frac{v_2 \ln a}{X_1} \quad (35)$$

for all $a$. Clearly, this equation is valid for arbitrary $a$ only if

$$\gamma_1 = \gamma_2 = 1 \quad (36)$$

so that we recover the logarithmic scaling. Thus we can conclude that data collapses for $\gamma_1 \neq 1$ or $\gamma_2 \neq 1$, if indeed realized, cannot be described within the framework of local scaling invariance presented here but rely on a different mechanism.

## 4. Conclusion

The main idea of Tang's concept of local scaling invariance is to consider continuously varying exponents which are constrained by a homomorphism. This homomorphism has the physical meaning that the exponents in a co-moving frame do not change under scaling transformations so that the compostion law (6) remains valid. The homomorphism can be expressed as a set of partial differential equations. We have shown that the solutions can be classified as follows:

- ordinary power-law scaling with constant exponents;
- logarithmic scaling, as observed numerically in [4, 8];
- scaling forms, where logarithmic and algebraic scaling are mixed;
- non-trivial scaling forms for more than two variables, classified by $m$.

Regarding possible applications we suggest that for any critical or self-similar system with logarithmically varying quantities these scaling forms are favourable candidates for a potential data collapse. Possible applications include phase transitions at the upper critical dimension, anomalous roughening transitions as well as disordered and self-organized critical systems. We would like to emphasize that this approach does not explain any type of data collapse, such as, for example, in equation (32), where logarithms are exponentiated. If such a data collapse is found, it has to be explained by a different theory.

Presently it is not yet clear whether and how these scaling forms are related to multiscaling and multifractals. Standard power-law scaling can only describe non-multifractal systems. We believe that the approach presented here can be useful to understand systems with multifractal properties. Both concepts are similar: in a multifractal different subsets of the system scale with different exponents, while in the present case different exponents are associated with different trajectories in the parameter space. These similarities indicate that both concepts may be closely related. In this context the approach by Jensen *et al* [12], who investigated multiscaling as a consequence of multifractality, may be useful.

As a far-reaching perspective we share Tang's point of view that the proposed concept of local scaling invariance needs to be substantiated by an appropriate renormalization group theory, by which the continuously varying exponents $\mathbf{v}(x_1, \ldots, x_n)$ can be computed. A contribution towards this direction can be found in [13].

## Acknowledgment

We would like to thank J Krug for sending us papers [4] and [9] and for pointing out that Tang's ideas are relevant for systems with logarithmic scaling.

**References**

[1] Binney J J, Dowrick N J, Fisher A J and Newman M E J 1992 *The Theory of Critical Phenomena* (Oxford: Clarendon)
[2] Barabàsi A L and Stanley H E 1995 *Fractal Concepts in Surface Growth* (Cambridge: Cambridge University Press)
[3] For an introduction to projective maps and homogeneous functions see e.g. Arnold V 1996 *Equations différentielles ordinaires* (Moscow: Mir)
    Arnold V 1996 *Chapitre suplémentaire de la théorie des équations différentielles ordinaires* (Moscow: Mir)
[4] Kadanoff L P, Nagel S R, Wu L and Zhou S M 1989 *Phys. Rev.* A **39** 6524
[5] Tebaldi C, De Menech M and Stella A L 1999 *Phys. Rev. Lett.* **83** 3952
[6] Meakin P, Coniglio A, Stanley H E and Witten T A 1986 *Phys. Rev.* A **34** 3325
[7] Wu X-Z, Kadanoff L, Libchaber A and Sano M 1990 *Phys. Rev. Lett.* **64** 2140
[8] Hinnemann B, Hinrichsen H and Wolf D E 2001 *Phys. Rev. Lett.* **87** 135701
    Hinnemann B, Hinrichsen H and Wolf D E *Phys. Rev.* E at press
[9] Tang C Scaling in avalanches and elsewhere *Preprint* NSF-ITP-89-118 (unpublished)
[10] Coniglio A and Marinaro M 1990 *Physica* A **163** 325–33
     Coniglio A and Marinaro M 1989 *Europhys. Lett.* **10** 575
[11] Landau L D and Lifshitz E M 1968 *Course of Theoretical Physics: Fluid Mechanics* vol 6 (Reading, MA: Addison-Wesley)
[12] Jensen M H, Paladin G and Vulpiani A 1991 *Phys. Rev. Lett.* **67** 208
[13] Stenull O and Jansen H K 2000 *Europhys. Lett.* **51** 539–45